\documentclass[prl,aps,twocolumn,superscriptaddress,preprintnumbers,nopacs,floatfix,amsmath,amssymb]{revtex4}

\usepackage{graphicx}
\usepackage{dcolumn}
\usepackage{bm}
\usepackage{amsmath}
\usepackage{graphicx}
\usepackage{amssymb}
\usepackage{mathrsfs}
\usepackage{bbm}
\usepackage{epsfig}

\newcommand{\be}{\begin{eqnarray}}
\newcommand{\ee}{\end{eqnarray}}

\begin{document}

%
%
%
\title{ The protein dynamical transition is a pseudogap changeover }

\author{Andrei Krokhotin}
\affiliation{Department of Physics and Astronomy, Uppsala University,
P.O. Box 803, S-75108, Uppsala, Sweden}
\author{Antti J. Niemi}
\affiliation{
Laboratoire de Mathematiques et Physique Theorique
CNRS UMR 6083, F\'ed\'eration Denis Poisson, Universit\'e de Tours,
Parc de Grandmont, F37200, Tours, France}
\affiliation{Department of Physics and Astronomy, Uppsala University,
P.O. Box 803, S-75108, Uppsala, Sweden}

\begin{abstract}
\noindent
The emergence of biochemical activities in a protein seem to commence with the onset of atomic mean-square 
displacements along the protein lattice.  The ensuing protein dynamical transition has
been discussed extensively in the literature, and often with conflicting conclusions. 
Here we clarify the phenomenon by establishing  a deep  connection between the 
dynamical  transition and the pseudogap state where high-temperature superconductivity comes to its end.
 For this we first show how to endow proteins with an order parameter akin the quasiparticle  
wave function in superconductors. We then present universality arguments to claim 
that the protein dynamical transition takes place in tandem with  a pseudogap transmutation. 
We confirm that  available experimental data fully supports our proposal.
\end{abstract}

\pacs{
87.15.H-, 87.64.kx,  87.10.Tf, 05.45.Yv, 36.20.Ey
}

\maketitle
The protein dynamical transition was first observed around 30 years ago in
myoglobin \cite{mos}-\cite{dow1}. It is now claimed  to be a common  property of all 
hydrated proteins \cite{ringe}-\cite{doster}. 
The transition takes place at temperatures that are somewhere between 180K-240K. 
Since this coincides with the  temperature range where proteins  quite universally 
start to display measurable  biochemical activities, it has been proposed that the  protein dynamical 
transition is concomitant  to the onset of life \cite{ringe}-\cite{doster}.  
But the detailed nature of the protein dynamical transition, even its very existence,  remains controversial
\cite{yun}, \cite{frau3}. The phenomenon, if it indeed exists, appears to closely  mimic the  properties
of the supercooled water  that surrounds the protein \cite{wat2}, \cite{wat1}:  At temperatures below 
$\sim$ 150K a flash-cooled  water  under ambient pressure is in an
amorphous, glassy state.  Between $\sim$ 150K-240K  there is  a  {\it no man's land} where the time scale for
crystallization is too short  to  be analyzed 
with present day techniques. But  there is evidence of  an ice-liquid coexistence  in this regime \cite{wat1}. 
At around 240K  water then enters a crystalline state with a homogeneous ice nucleation that persists 
until  the liquid phase takes over at 273K. 

To a large extent protein folding is known to be driven by a combination of hydrophobic and 
hydrophilic effects. Therefore the  properties of a hydrated protein should  mirror 
those of the adjacent water. Indeed, there is  a wide consensus that the protein 
dynamical transition is, likewise a glass transition,  driven  by two intertwining 
processes that are controlled by a combination  of the different properties that  water has in the 
thin hydration shell  and in the  surrounding bulk.  At an atomary level the two processes are as 
follows \cite{ringe}-\cite{doster}:  At temperatures 
below $\sim $ 180K the covalent protein lattice is  in a state where the individual 
atoms are predominantly subject to local  thermal fluctuations. These 
fluctuations can be interpreted as  simple harmonic vibrations of  atoms around their lattice 
equilibrium positions. The Debye-Waller relation   connects the atomic mean square 
displacements to the experimental B-factors, that grow 
linearly at  low temperatures 
\begin{equation}
<x^2>_T  \ \approx  \frac{B}{8\pi^2}   \simeq a\cdot T + b
\label{B}
\end{equation}
When the protein dynamical transition takes place, atomic displacements  begin to correlate
and start covering much larger length scales. Various collective motions such  as  fluctuations 
between different  macromolecular sub-states  make the scene, take over and  rule the internal dynamics 
of the protein lattice all the way to physiological temperatures. This becomes reflected in
the B-factors that should now  display a more rapid, maybe even exponential increase  as a function of
temperature.

The conventional approach to the protein dynamical transition is very much concentrated on understanding
the fine structure 
and detailed properties of supercooled water  \cite{ringe}-\cite{doster}.  Here we propose a radical departure 
from the paradigm way. Our approach is based on the theoretical concept of {\it universality} that already has  a
firm  basis in the description of phase properties of matter in terms of a relatively small number of relevant
and marginal interactions \cite{zin}. We employ  a combination of universality arguments  with  experimental data analysis to conclude  
that the protein dynamical transition must be a pseudogap changeover.  We argue that the pertinent 
pseudogap  state is the one that precedes the $\Theta$-point phase transition where proteins 
become  denatured  and depart from their  biologically active collapsed phase.  Furthermore, since our arguments
are entirely based on the concept of universality, the transition can not be specific to proteins and water  only, it 
is a more general property of polymers in bad solvents.

A pseudogap state was originally  introduced to explain aspects of high temperature superconductors \cite{pseudo1}, \cite{pseudo2}.
Subsequently it has been found experimentally  even in  ferromagnetic metals 
\cite{naturemet}. Theoretically,   a pseudogap has been proposed to be a participant  both in color superconductors and in
the chiral phase transition of QCD  \cite{babaev}, \cite{zarembo}.
At a theoretical level and  in its simplest form,  the pseudogap state is described by a single complex 
order parameter $\phi = \rho \exp\{i\theta\}$.   In the case of superconductivity  
this could be the order parameter for  the Cooper pairing  of electrons. In the symmetric phase where there is
no superconductivity, the free energy has its minimum when the modulus $<|\phi |> = <\rho> $ vanishes. 
The symmetry becomes broken when there is condensation of $\phi$ and  the free energy is minimized 
by a non-vanishing value of the modulus $< \rho> $. 
In the case of  a superconductor this takes place when electrons combine 
into Cooper pairs, the hallmark  of a superconducting phase. 
The pseudogap state is a refinement of this  phase structure  into the case where  $<\phi> $ vanishes even though 
 $<\rho>$ retains its  non-vanishing symmetry breaking value \cite{pseudo1}. This  occurs in the presence of 
a strong phase decoherence,
\begin{equation}
<\exp\{ i\theta \}> = 0
\label{psgap}
\end{equation}
Consequently we can have $<\phi> = 0$ even though the nonvanishing gap $<\rho>$ persists. The  pseudogap state is  like
a symmetric phase  precursor state in the  broken symmetry phase, a prelude to the  transition that takes place 
when the gap $<\rho>$ eventually vanishes.

In the case of proteins,  more generally of polymers,  the conventional order parameter is the
compactness index $\nu$ that describes how the  radius of gyration $R_g$  scales in the 
limit where the  number $N$ of residues becomes very large \cite{degennes},
\begin{equation}
R_g \ = \ \sqrt{ \frac{1}{2N^2}  \sum_{i,j} ( {\bf r}_i 
- {\bf r}_j )^2 } \  \approx \  R_0 N^{\nu} 
\label{nu}
\end{equation} 
Here ${\bf r}_i$ ($i=1,2,...,N$) are the locations of the central C$_\alpha$ carbons and $R_0$ is a 
protein specific but $N$ independent pre-factor.   
Canonically, the compactness index can assume only  three different  values. The mean-field theory  value 
$\nu = 1/3$ corresponds to the biologically active
collapsed phase where proteins are in a space filling conformation. This commonly
occurs at low temperatures or in a bad solvent environment. 
When  $\nu = 1/2$ we have a fully flexible chain, this corresponds to the $\Theta$-point. 
Finally, $\nu = 3/5$ is the mean-field Flory value for the universality class  of self-avoiding random walk that
describes a polymer either at a very high temperature or
in a very good solvent environment.  
This conventional approach to protein phase structure does not have the flexibility  to describe a  pseudogap state,
so we now propose a refinement.  We start  by introducing the local  backbone bond  and torsion  
angles. We utilize the C$_\alpha$ coordinates to define the backbone unit tangent ($\bf t$) and binormal ($\bf b$) vectors
\begin{equation}
{\bf t}_i = \frac{ {\bf r}_{i+1} - {\bf r}_i }{ | {\bf r}_{i+1} - {\bf r}_i |}
 \  \ \ \& \ \ \ {\bf b}_i = \frac{ {\bf t}_{i-1} \times {\bf t}_i }{| {\bf t}_{i-1} \times {\bf t}_i|}
\label{tb}
\end{equation}
Then 
\begin{equation}
\kappa_{ i} = \arccos ( {\bf t}_{i+1} \cdot {\bf t}_i ) \  \ \ \& \ \ \  \tau_i 
= \arccos ({\bf b}_{i+1} \cdot {\bf b}_i) 
\label{tors}
\end{equation}
are the discrete bond and torsion angles, respectively. Note that  these definitions involve only  
the positions of the C$_\alpha$ carbons. Consequently each quantity is inherently geometrical
and  we can introduce  the similarly geometrical order parameter
\begin{equation}
\psi_i  = \kappa_i \cdot \exp\{ i\tau_i\}
\label{psi}
\end{equation}
In fact, this is {\it exactly} the Hasimoto identification of the wave function in nonlinear Sch\"odinger equation \cite{hasi},    
in its discrete version.  As a complex variable the order parameter (\ref{psi}) also has the requisite structure to 
describe  a pseudogap. We now proceed to argue that in a collapsed protein the pseudogap state also occurs, and that it
is  a precursor state to
the $\Theta$-point phase transition.

In \cite{xubiao}  it has been shown  that most proteins in Protein Data Bank, over 90$\%$ of them, 
can be described with experimental B-factor accuracy in terms of soliton solutions to 
the following variant of the discrete nonlinear Schr\"odinger equation \cite{nora}
\[
E = - \sum\limits_{i=1}^{N-1}  2\, \kappa_{i+1} \kappa_i  + \sum\limits_{i=1}^N
\biggl\{  2 \kappa_i^2 + q\cdot (\kappa_i^2 - m^2)^2  
\]
\begin{equation}
\left. + \frac{d_\tau}{2} \, \kappa_i^2 \tau_i^2  -  b_\tau \kappa_i^2  \tau_i - a_\tau  \tau_i   +  \frac{c_\tau}{2}  \tau^2_i 
\right\} 
\label{E1}
\end{equation}
The first sum together with the three first terms in the second sum comprise  exactly   the 
energy of  the standard discrete nonlinear Schr\"odinger equation  (DNLS) when expressed in terms of the  
Hasimoto  variable (\ref{psi}).  The fourth ($b_\tau$) and fifth ($a_\tau$) terms are the two
conserved quantities  that precede the energy  in the DNLS hierarchy,  the momentum and the helicity  \cite{fadde}.   
The last ($c_\tau$) term is the  Proca mass term, we include it only  for completeness.

The energy function (\ref{E1}) does not purport to explain the 
details of the atomary level mechanisms that give rise to protein folding. Rather, it allows  us to examine the properties of a folded protein 
backbone in terms of universal physical arguments, much like an  effective Landau-Ginzburg model describes superconductivity.  
In fact,  we can further develop the analogy between (\ref{psi}) and Cooper pairs,
by noting that (\ref{E1})  has the functional form of the discretized  Landau-Ginzburg free energy 
in the supercurrent variables \cite{degennes2}: In the continuum limit the first two terms combine into the 
derivative of $\kappa(x)$ that plays the r\^ole of  Cooper pair density. The third term is the standard 
symmetry breaking potential and the fourth term has its origin in spontaneous symmetry breaking that
leads to the notorious Meissner effect \cite{degennes2}.  In addition we have included in (\ref{E1}) {\it exactly} 
all those terms  that are consistent with general principles of universality and gauge invariance.  

If we use the $\tau_i$ equation of motion to eliminate the torsion angles,  the potential  for the bond angles becomes
\begin{equation}
E_{pot} =  \frac{ad+bc}{d (c + d \kappa^2) } +  2\left(1-q m^2 - \frac{b^2}{4d} \right) \kappa^2
+ q \kappa^4
\label{U}
\end{equation}
The first term relates  to the potential energy in a Calogero-Moser system,   notorious for its r\^ole in describing fractional
statistics.  Here it is slightly deformed by the Proca mass. The second and third terms have the conventional form of a 
symmetry breaking double-well potential.  Depending on the parameter values we may  be either in the symmetric
$\kappa_i = 0$  phase or in the broken symmetry phase where $\kappa_i $ acquires a 
non-vanishing ground state value. In the second case the local energy minimum states correspond to regular protein secondary 
structures such as $\alpha$-helix or $\beta$-strand and protein loops are like domain wall configurations that interpolate between 
the different local  minima \cite{cherno}. Explicit computations confirm \cite{martin}  that the symmetry breaking 
$\nu \approx 1/3$ low temperature phase where long-range correlations rule, becomes converted 
into a symmetry restored  $\nu \approx 1/2$ phase at higher temperatures.  
In particular, the symmetry restoration is a phase transition that takes place when temperature reaches the $\Theta$-point value
that separates the low temperature collapsed phase from the higher temperature ideal chain phase  \cite{martin}.  

Unlike in the case of conventional metallic superconductors where the supercurrent has stiffness that leaves no room for a pseudogap, the torsion 
angles $\tau_i$ in (\ref{E1}) are only subject to ultra-local couplings and as such they are highly flexible. 
Thus  it is conceivable that as temperature starts increasing
in the low temperature phase,  there will also be an increase in phase decoherence that is measured by  $<\exp\{ i \tau\} >$. Depending on
the dynamical details,  the energy function (\ref{E1})  may then describe a pseudogap state at  temperatures
below  the $\Theta$-point value. 
But as it stands, the present energy function can not fully model the pseudogap transition in proteins. This is because
the model does not include the side-chains  with their numerous rotamers  $\chi_i$. The rotamers
give us plenty of opportunities to introduce many additional  pseudogap detecting 
order parameters $ <\exp\{ i \chi_i\}> $ in proteins: 

At low temperatures protein is in a crystalline state and the rotamers assume very definite values
that are catalogued in various libraries. As temperature raises  fluctuations appear, but these are
damped by various steric constraints that strongly limit  the ability of  individual atoms to move.  In proteins the steric constraints are  more lenient  to the outlying side-chain rotamers  than to the backbone torsion angles:  With fever  covalent neighbors,  there is more room to fluctuate. 
Consequently any protein pseudogap transition should  become primarily 
visible in an increased phase de-coherence in the
outlying  side-chain rotamers. But  even though steric constraints are more 
lenient to side-chain rotamers, the identification of a protein 
pseudogap is much more delicate than the identification of a pseudogap state in materials such as  
high-temperature superconductors  where no steric constraints exist. 
Despite being numerous, the  protein side-chain rotamers can never assume unhindered, mutually fully uncorrelated
 values. Consequently protein rotamer angles can never display a  full, unrestrained phase de-coherence the way 
 how it  is laid out by a high temperature superconductor. Rather we expect there to be  a  {\it pseudogap changeover} that makes 
its presence known in a marked, rapid  relative increase in the side-chain rotamer fluctuations as the temperature climbs up towards
the $\Theta$-point value.

Presently, there are no direct experimental measurements of  temperature dependence in the rotamer pseudogap order parameters
$ <\exp\{ i \chi_i\}> $.  We propose that such experiments could be performed.
At the moment our ability to investigate the inception of a
pseudogap  is limited  to the analysis of increased temperature dependence in  the experimental side-chain 
B-factors that indirectly measure the $ <\exp\{ i \chi_i\}> $ fluctuations. We have screened all data available in Protein Data Bank,
but mutually compatible good quality PDB data on B-factors is  sparse. Structures have  been determined without
standardized crystalline environments, using different refinement procedures, and with apparently different practices
for determining the B-factors. This makes  it very hard for us to compare data that is collected in different 
experiments and at different  temperatures.  Consequently our ability to reliably investigate the relation between 
a pseudogap changeover and  the protein dynamical transition 
is limited to the re-analysis of data  that has been collected in only two experiments,
with  crambin \cite{crambin} and ribonuclease A \cite{petsko}.  
They both measure B-factors at different temperatures, consistently with the same protein crystals. 
We have re-analyzed the data and find that both experiments {\it fully} support  our proposal that the protein dynamical
transition is a pseudogap changeover.

\begin{figure}[!hbtp]
  \begin{center}
    \resizebox{8.5cm}{!}{\includegraphics[]{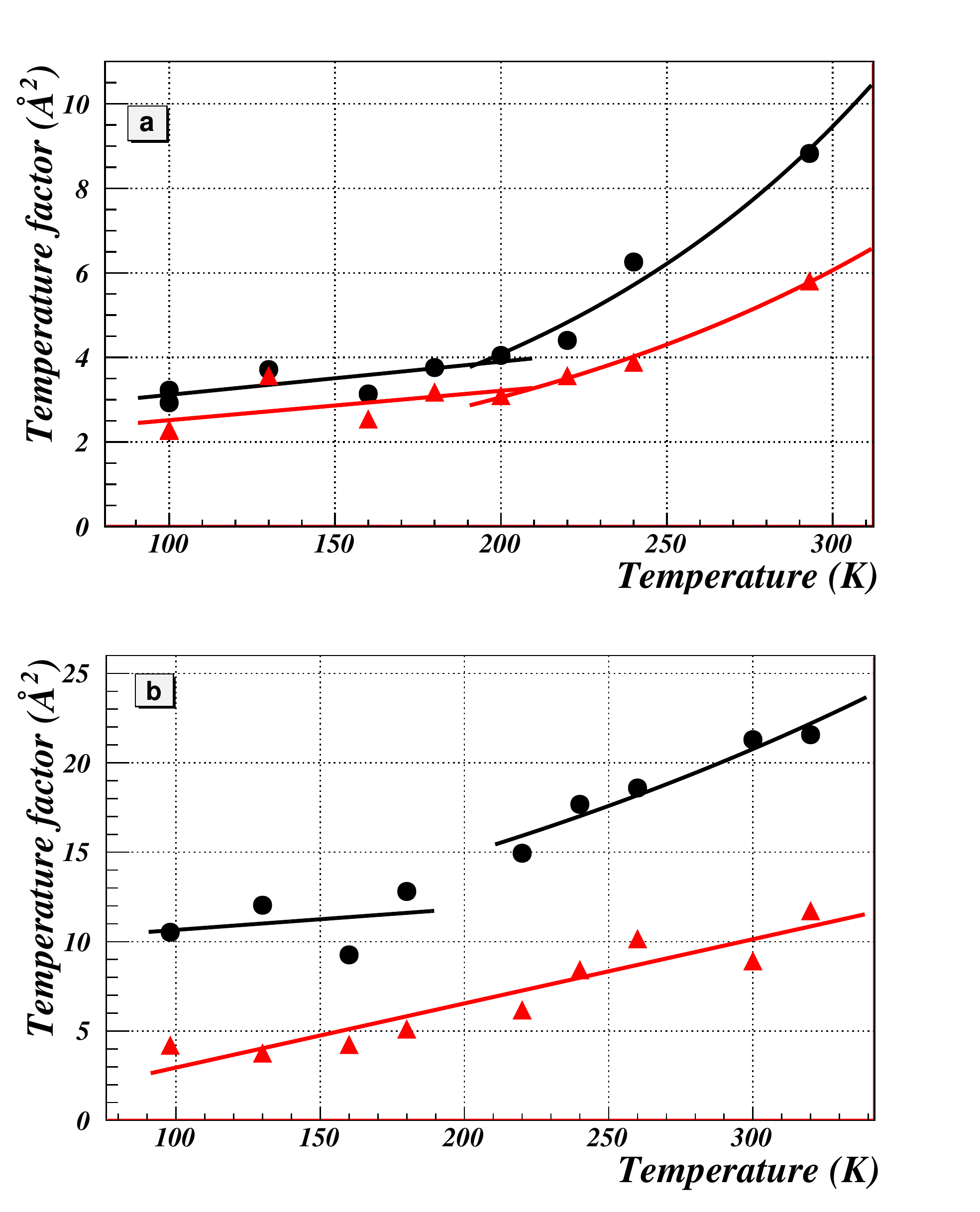}}
    \caption{ a) The B-factors of crambin  from \cite{crambin} as a function of temperature, separately for backbone C$_\alpha$ (red) and side-chain atoms (black) but excluding the C$_\beta$. b) Same as a)  but for ribonuclease A, with data from  \cite{petsko}.  In both cases the side-chains display a clear transition and  nonlinear $T$ dependence, while for the C$_\alpha$ even the linear fit (\ref{B}) is feasible.}
    \label{fig:simple}
  \end{center}
\end{figure}
In Figure 1 we display the B-factors of these two experiments as a function of temperature, 
separately for the backbone C$_\alpha$ carbons  and for the side-chain atoms.
From  the side-chains we exclude the C$_\beta$ atoms, they are slaved to the sp3 hybridized backbone C$_\alpha$ atoms
both by covalent bonds and by existing refinement procedures.  In the case of crambin we observe a clearly visible  qualitative 
change and a definite enhancement in the temperature dependence of the side-chain B-factors in comparison
 to the backbone C$_\alpha$ atoms. There is  also a 
 transition from a linear to a non-linear behavior. The change commences  with the  protein dynamical 
 transition regime. For ribonuclease A the results are similar, 
 but the effect is less profound.  We remark that in both cases the B-factors of the C$_\alpha$ atoms  truly 
 deviate only slightly  (if at all in the case of ribonuclease A) from linearity in $T$. This is consistent with \cite{xub}, where
 no sign of a protein dynamical transition was observed at the level of  the backbone C$_\alpha$ carbons.
  
 In summary, the protein dynamical transition and the pseudogap state in high temperature superconductors are two {\it a priori}
 totally different physical phenomena. However, here we have proposed that they are intimately related by asserting  that
 the protein dynamical  transition is a pseudogap changeover. Our argumentation is based both on
 theoretical  observations that stem from the concept of universality, and analysis of all presently available 
 experimental data.   We propose that experiments could be performed to directly detect the pseudogap
 transition, in particular since the possibility that a pseudogap converts a static protein crystal into a 
 biologically active nano-engine  could have really far reaching consequences in the search for a physical 
origin of  life.

\vskip 0.3cm
A.J.N. thanks H. Frauenfelder for communications, and for providing a copy of \cite{frau3}.

\vskip 2.6cm


\begin{thebibliography}{99}

\bibitem{mos}  F. Parak,  E.N. Frolov, R.L. M\"ossbauer and V.I. Goldanskii,
J. Mol. Biol. {\bf 145} 825
(1981)

\bibitem{frau} R.H. Austin, K.W. Beeson, L. Eisenstein, H. Frauenfelder and I.C.
Gunsalus, 
Biochemistry {\bf 14}  5355
(1975).

\bibitem{dow1} W. Doster, S. Cusack and W. Petry,
Nature {\bf 337} 754
(1989).

\bibitem{ringe}  D. Ringe and G.A.  Petsko, 
Biophys. Chem. {\bf 105}  667
(2003)

\bibitem{frau2} H. Frauenfelder   {\it et.al.}
PNAS {\bf 106} 5129
(2009)

\bibitem{doster} W. Doster, 
Biochim. Biophys. Acta {\bf 1804}  3
(2010)

\bibitem{yun} Y. He, P.I. Ku, J.R. Knab, J.Y. Chen and A.G. Markelz, 
Phys. Rev. Lett. {\bf 101} 178103
(2008)  

\bibitem{frau3} R.D. Young, H.Frauenfelder and P.W. Fenimore, preprint LA-UR 11-00161

\bibitem{wat2} A. Mishima and H.E. Stanley,
Nature {\bf 396}  329
(1998)

\bibitem{wat1} E.B. Moore and V. Molineroa, 
J. Chem. Phys. {\bf 132}  244504
(2010)

\bibitem{zin} J. Zinn-Justin,  {\it Quantum field theory and critical phenomena} (Clarendon Press, Oxford,  2002)

\bibitem{pseudo1} V.J. Emery and S.A. Kivelson, 
Nature 374, 434
(1995).

\bibitem{pseudo2}  J. Corson, R. Mallozzi, J. Orenstein, J.N. Eckstein and I. Bozovic, 
Nature {\bf 398}, 221
(1999)

\bibitem{naturemet} N. Mannella {\it et.al.} 
Nature 438, 474
(2005)


\bibitem{babaev} H. Kleinert and E. Babaev, 
Phys. Lett. {\bf B438} 311
(1998)


\bibitem{zarembo} K. Zarembo, 
JETP Letters {\bf 75} 59
(2002)


\bibitem{degennes} P.G. De Gennes,  {\it Scaling Concepts in Polymer Physics} (Cornell 
University Press, Ithaca, 1979)

\bibitem{hasi} H. Hasimoto, 
J. Fluid Mech. {\bf 51} 477
(1972)

\bibitem{xubiao} A. Krokhotin,  A.J. Niemi and X. Peng, e-print 	arXiv:1109.3903v1 [physics.bio-ph]

\bibitem{nora}  N. Molkenthin,  S. Hu and A.J.   Niemi,  
Phys. Rev. Lett. {\bf 106}  078102
(2011) 

\bibitem{fadde} L.D. Faddeev and L.A. Takhtajan,   {\it Hamiltonian methods in the theory of solitons}
(Springer Verlag, Berlin, 1987) 
	

\bibitem{degennes2}  P.G. De Gennes,  {\it Superconductivity 
of Metals and Alloys} (Westfield Press, New York 1995)

\bibitem{cherno} M. Chernodub, S. Hu and A.J. Niemi, 
Phys. Rev. {\bf E82} 011916
(2010)


\bibitem{martin}  M. Chernodub, M. Lundgren and A.J. Niemi,
Phys. Rev. {\bf E83} 011126
(2011)


 \bibitem{crambin}  M.M. Teeter, A. Yamano, B. Stec and U. Mohanty, 
  Relation to protein function 
  PNAS {\bf 98} 11242
  (2001)

 \bibitem{petsko} R.F. Tilton, J.C.  Dewan and G.A.  Petsko, 
 Biochemistry {\bf 31}  2469
 (1992)
 
\bibitem{xub} A. Krokhotin, A.J. Niemi and X. Peng, 
(to appear)

\end{thebibliography}
\end{document}